\documentclass[useAMS,usenatbib]{mn2e}
\usepackage{times,graphicx}
\newcommand{\apj}{ApJ}                 
\newcommand{\mnras}{MNRAS}             \newcommand{\aap}{A\&A}

               \newcommand{\physrep}{Physics Reports}
\newcommand{\dd}{{\rm d}}
\newcommand{\vc}[1]{\textbf{\emph #1}}
\title[Equilibrium state of self-gravitating systems]
{Saddle-point entropy states of equilibrated self-gravitating systems}
\author[P. He and D.-B. Kang]{Ping He\thanks{Email: hep@itp.ac.cn} and Dong-Biao Kang \\
Key Laboratory of Frontiers in Theoretical Physics, Institute of Theoretical Physics, Chinese Academy of Sciences, Beijing 100190, China}
\date{\today}
\begin{document}\maketitle
\label{firstpage}

\begin{abstract}
In this Letter, we investigate the stability of the statistical equilibrium of spherically symmetric collisionless self-gravitating systems. By calculating the second variation of the entropy, we find that perturbations of the relevant physical quantities should be classified as long- and short-range perturbations, which correspond to the long- and short-range relaxation mechanisms, respectively. We show that the statistical equilibrium states of self-gravitating systems are neither maximum nor minimum, but complex saddle-point entropy states, and hence differ greatly from the case of ideal gas. Violent relaxation should be divided into two phases. The first phase is the entropy-production phase, while the second phase is the entropy-decreasing phase. We speculate that the second-phase violent relaxation may just be the long-wave Landau damping, which would work together with short-range relaxations to keep the system equilibrated around the saddle-point entropy states.
\end{abstract}

\section{Introduction}
\label{sec:intro}

It has been realized that the conventional methods of statistical mechanics of short-range interaction systems cannot be directly carried out to study long-range self-gravitating systems. Hence, it is necessary to return to the starting point of statistical mechanics and develop special techniques to handle the long-range nature of gravity \citep{padmanabhan90}. In our earlier two works \citep{hep10, kang11}, we performed preliminary investigations of statistical mechanics of collisionless self-gravitating systems and found that (1) both the concept of entropy and entropy principle are still valid for self-gravitating systems; (2) entropy is additive and hence extensive; (3) the equilibrium states also consist of mechanical equilibria; and (4) systems' local and global equilibria are different.

Based on these findings, in \citet{hep11a}, we formulate a systematic theoretical framework of the statistical mechanics of spherically symmetric collisionless self-gravitating systems, with innovative approaches that differ significantly from the conventional statistical mechanics of short-range interaction systems. First, we demonstrate that the equilibrium states of self-gravitating systems consist of both mechanical and statistical equilibria, the former being characterized by a series of velocity-moment equations and the latter by the statistical equilibrium equations, which should be derived from entropy principle. The velocity-moment equations for spherical systems of all orders are derived from the steady-state collisionless Boltzmann equation. Then, we point out that ergodicity is invalid globally for self-gravitating systems, but can still be reestablished locally if gravitating particles are treated as indistinguishable. Based on the local ergodicity, by using Fermi-Dirac statistics, with the weakly-degenerate condition and the spatial independency of the microstates, we re-derive the Boltzmann-Gibbs entropy, which should be exactly the correct entropy form for collisionless self-gravitating systems. Apart from the usual constraints of given mass and energy, we demonstrate that the series of moment equations must be included as additional constraints on the system's entropy functional when performing the variational calculus, which is an extension to the original prescription by \citet{white87}.

In that work, we left behind a question whether the statistical equilibria of gravitating systems, worked out with our approach, are stable or not. In this Letter, we investigate the stability of the statistical equilibria of spherically symmetric systems by calculating second variation of entropy. The Letter is organized as follows. In Section~\ref{sec:solution}, we present solutions with a truncated distribution function (DF), derived in \citet{hep11a}. In Section~\ref{sec:stability}, we calculate second variations of both ideal gas and gravitating systems to make a close comparison between these two systems. We present discussions and conclusions in Section~\ref{sec:concl}.

\section{Solutions based on the first variation}
\label{sec:solution}

We briefly summarize the basic equations derived in \citet{hep11a}, on the basis of the first variation of entropy, with the truncated DF up to the second order of velocity expansion.

The coarse-grained DF can be formally expressed as
\begin{equation}
\label{eq:lmpd2}
F(\vc{x}, \vc{v}) = \exp\big(-\sum_{k,m,n} \lambda_{k,m,n} (\vc{x}) v^k_1 v^m_2 v^n_3 \big),
\end{equation}
where $\lambda_{k,m,n} (\vc{x})$ correspond to the formal Taylor expansion coefficients of $\ln F(\vc{x}, \vc{v})$ with respect to $\vc{v}$. These undetermined coefficients should be determined by the joint mechanical and statistical equilibrium equations. At this moment, we do not know how to evaluate the entropy and derive the equations of statistical equilibria with this complete DF and hence some truncation to the DF is necessary. We consider that, with this truncation, although the results are not accurate, they are still valuable and heuristic.

For the spherically symmetric systems, the truncated DF $F(\vc{x}, \vc{v})$ of equation~(\ref{eq:lmpd2}) to the second order can be expressed as
\begin{equation}
\label{eq:3df}
F(r, v_r, v_{\theta}, v_{\phi}) = \frac{\rho}{(2\pi)^{\frac{3}{2}}\sigma_r \sigma_{\theta} \sigma_{\phi}} {\rm exp} \big( -\frac{v^2_r}{2 \sigma^2_r} - \frac{v^2_{\theta}}{2\sigma^2_{\theta}} - \frac{v^2_{\phi}}{2 \sigma^2_{\phi}}\big),
\end{equation}
in which the density $\rho$, the velocity dispersions $\sigma^2_r$, $\sigma^2_{\theta}$ and $\sigma^2_{\phi}$ are all functions of $r$. As a result, the total (Boltzmann-Gibbs) entropy of the system with this truncated DF is
\begin{displaymath}
S_{\rm BG} = -\int F \ln F \dd^3\vc{x} \dd^3\vc{v} = \int^{\infty}_{0} 4\pi r^2 \rho \left[ -\ln\big(\frac{\rho}{\sigma_r \sigma_{\theta} \sigma_{\phi}}\big)\right.
\end{displaymath}
\begin{equation}
\label{eq:bgt}
{\hskip 0mm} + \left.\frac{3}{2} (1 + \ln 2\pi)\right] \dd r = \int^{\infty}_{0} 4\pi r^2 \rho \ln\big(\frac{p^{1/2}_r p^{1/2}_{\theta} p^{1/2}_{\phi}}{\rho^{5/2}}\big)\dd r,
\end{equation}
where in the last equality we drop the unimportant constant term, $\frac{3}{2} (1 + \ln 2\pi)$, and $p_r \equiv \rho \sigma^2_r$, $p_{\theta} \equiv \rho \sigma^2_{\theta}$ and $p_{\phi} \equiv \rho \sigma^2_{\phi}$.

We define the kinetic energies that are contained in the $r$-sphere of the system in three orthogonal directions as
\begin{eqnarray}
\label{eq:xyze}
k_{\theta}(r) & = & \frac{1}{2}\int^r_0 4\pi r'^2 p_{\theta}(r')\dd r', \nonumber \\
k_{\phi}(r) & = & \frac{1}{2}\int^r_0 4\pi r'^2 p_{\phi}(r')\dd r', \nonumber \\
k_{r}(r) & = & \frac{1}{2}\int^r_0 4\pi r'^2 p_r(r')\dd r'.
\end{eqnarray}
Thus, the total kinetic energy contained in the sphere is
\begin{equation}
\label{eq:ke}
k_{\rm t}(r) = k_{\theta}(r) + k_{\phi}(r) + k_{r}(r)
\end{equation}
and the mass function $m(r)$ is
\begin{equation}
\label{eq:mf}
m(r) = \int^r_0 4 \pi r'^2 \rho(r')\dd r',
\end{equation}
and $u(r)$ is the potential energy of a spherical gravitating system:
\begin{eqnarray}
\label{eq:ue}
u(r)& = & -\int_{\Omega_r}\frac{G \rho(\vc{r}') \rho(\vc{r}'')}{2 | \vc{r}' - \vc{r}'' |} \dd^3 \vc{r}' \dd^3 \vc{r}'' \nonumber \\
    & = & -4 {\pi} G \int^r_{0}m(r')\rho(r')r' \dd r',
\end{eqnarray}
where the integral is restricted to the volume of the $r$-sphere, $\Omega_r$. With these variable definitions, the total mass $M_{\rm T}$, total kinetic energy $E_{\rm K}$, and total potential energy $E_{\rm V}$ of the system are, respectively,
\begin{equation}
\label{eq:consv}
M_{\rm T} = m(r)|_{r \rightarrow \infty}, {\hskip 0.75mm} E_{\rm K} = k_{\rm t}(r) |_{r \rightarrow \infty} {\hskip 0.75mm} {\rm and} {\hskip 0.75mm} E_{\rm V} = u(r)|_{r \rightarrow \infty}.
\end{equation}
From equations~(\ref{eq:xyze}) and (\ref{eq:mf}), we perform the variable transformation by calculating the first derivatives as
\begin{equation}
\label{eq:vd3}
\rho=\frac{m'}{4\pi r^2}, {\hskip 0.35cm} p_{\theta}=\frac{k'_{\theta}}{2\pi r^2}, {\hskip 0.35cm} p_{\phi}=\frac{k'_{\phi}}{2\pi r^2}, {\hskip 0.35cm} p_{r}=\frac{k'_r}{2\pi r^2},
\end{equation}
in which the superscript `prime' denotes the first derivative with respect to $r$. With these ready, the total entropy of the system from equation~(\ref{eq:bgt}) is
\begin{eqnarray}
\label{eq:tots2}
S_{\rm t} & = & \int^{\infty}_0 H_1(r, m, m', k_{\rm t}, k'_{\rm t}, k_{\theta}, k'_{\theta}, k_r, k'_r) \dd r \nonumber \\
& = & \int^{\infty}_0 \left\{ m'\ln \left[ \frac{k'^{1/2}_{\theta} k'^{1/2}_r (k'_{\rm t} - k'_{\theta} - k'_r)^{1/2}} {m'^{5/2}} \right]\right. \nonumber \\
& & + \left. m' \ln\big(2^{7/2} \pi r^2 \big) \right\} \dd r,
\end{eqnarray}
where $H_1$ denotes the integrand, the terms enclosed in the braces.

As demonstrated in \citet{hep11a}, the equilibrium states of self-gravitating systems consist of both mechanical and statistical equilibria, the former being characterized by a series of velocity-moment equations. They are derived from the steady-state collisionless Boltzmann equation and should also be included as additional constraints on the entropy functional when performing the variational calculus, besides the usual constraints of mass and energy conservation. The second-order moment equation is just the familiar Jeans equation:
\begin{equation}
\label{eq:2ndoe}
\frac{\dd}{\dd r}(\rho\overline{v^2_r}) + \frac{2\rho}{r}( \overline{v^2_r} - \overline{v^2_t}) = \frac{\dd}{\dd r}(\rho\overline{v^2_r}) + \frac{2\beta}{r} \rho\overline{v^2_r} = - \rho \frac{\dd \Phi}{\dd r},
\end{equation}
where $v_t$ denotes either $v_{\theta}$ or $v_{\phi}$, and the barred quantities indicate the corresponding velocity moments. $\beta$ is the usual velocity anisotropy parameter, defined as $\beta = 1 - \overline{v^2_t}/\overline{v^2_r}$.

The Jeans equation~(\ref{eq:2ndoe}) is equivalent to the following virialization form \citep*[e.g.][p.235]{mo10}:
\begin{equation}
\label{eq:vir1}
2k_{\rm t}(r) + u(r)- 2 r k'_r(r) = 0.
\end{equation}
We will use this relation, instead of the original Jeans equation~(\ref{eq:2ndoe}), as the constraint of the mechanical equilibrium. Additionally, from equation~(\ref{eq:ue}), we have
\begin{equation}
\label{eq:ub1}
u' = -4 {\pi} G m \rho r = -\frac{G m m'}{r},
\end{equation}
where the superscript `prime' again indicates the first derivative with respect to the radius $r$. The mathematical relation between $u$ and $m$ of equation~(\ref{eq:ub1}) provides another additional constraint.

With all these constraints considered, the total constrained entropy of the system is
\begin{displaymath}
S_{\rm t, c} =  \int^{\infty}_0 H_2(r, m, m', k_{\rm t}, k'_{\rm t}, k_{\theta}, k'_{\theta}, k_r, k'_r, u, u')\dd r
\end{displaymath}
\begin{displaymath}
{\hskip 7.5mm} = \int^{\infty}_0 \left\{m'\ln\left[\frac{k'^{1/2}_{\theta} k'^{1/2}_r (k'_{\rm t}-k'_{\theta} - k'_r )^{1/2}}{m'^{5/2}} \right]\right.
\end{displaymath}
\begin{displaymath}
{\hskip 10.9mm} + m' \ln(2^{7/2} \pi r^2) + f_1(r)(2k_{\rm t} + u - 2 r k'_r)
\end{displaymath}
\begin{equation}
\label{eq:totcs}
{\hskip 10.9mm} + \left. f_2(r)(u' + \frac{G m m'}{r}) \right\} \dd r,
\end{equation}
where $H_2$ denotes the integrand enclosed in the braces, and $f_1$ and $f_2$ are two Lagrangian multipliers. By using the variable transformations of equations~(\ref{eq:ke}) and (\ref{eq:mf}), the mass and energy conservation constraints for the variational calculus are converted to satisfying the fixed end-point conditions of equation~(\ref{eq:consv}). For spherical systems, $p_{\theta}(r)=p_{\phi}(r)$, so $k_{\theta}(r) = k_{\phi}(r)$. Then perform the standard variational calculus, $\delta S_{\rm t,c} = 0$, with respect to the variable pairs $(m, m'), (k_{\rm t}, k'_{\rm t}), (k_r, k'_r)$, and $(u, u')$, we obtain the following two equations:
\begin{displaymath}
\frac{\dd\ln\rho}{\dd r}-\frac{\dd\ln p_t}{\dd r}-\frac{p_t}{r p_r}+\frac{1}{r}=0,
\end{displaymath}
\begin{equation}
\label{eq:ese}
\frac{3}{2}\frac{\dd \ln\rho}{\dd r} - \frac{1}{p_t} \frac{\dd p_r}{\dd r}-\frac{2 p_r}{r p_t} + \frac{2}{r} = \lambda\frac{G m}{r^2},
\end{equation}
where $p_t$ denotes either $p_{\theta}$ or $p_{\phi}$, with the two Lagrangian multipliers as
\begin{equation}
\label{eq:lmf}
f_1 = \frac{\rho}{2 r p_r} - \frac{\rho}{2 r p_t}, {\hskip 10mm} f_2=\frac{\rho}{2p_t} - \lambda,
\end{equation}
where $\lambda$ is an integration constant and should be always negative \citep[see][]{hep10, kang11}.

\section{Stability based on the second variation}
\label{sec:stability}

The above equations are just the lowest-order approximation, derived from the truncated DF of equation~(\ref{eq:3df}), but, as mentioned previously, we may still acquire useful knowledge with this approximated treatment. We investigate the stability of statistical equilibria of the self-gravitating systems by analysing the second variations of entropy.

\subsection{Ideal gas}
\label{sec:idg}

We put an ideal gas in a spherical container and perform the variational calculus in a similar way to that for spherical self-gravitating systems to make a close comparison between the two systems.

Since the Boltzmann-Gibbs entropy is suitable for short-range interaction systems, so the entropy of equation~(\ref{eq:tots2}) can also be applied to ideal gas. We perform the first variational calculus with respect to all the variable pairs $(m, m')$, $(k_{\rm t}, k'_{\rm t})$, $(k_{\theta}, k'_{\theta})$ and $(k_r, k'_r)$, and obtain $\rho = \rho_0$, and $p_{\theta} = p_{\phi} = p_r = p_0$, in which $\rho_0$ and $p_0$ are two constants. We then calculate the second variation and obtain
\begin{displaymath}
\delta^2 S_{\rm t} =  \frac{1}{2} \int^{\infty}_0 [A_1(\delta m)^2 + B_1(\delta m')^2 + A_2(\delta k_{\rm t})^2 + B_2(\delta k'_{\rm t})^2
\end{displaymath}
\begin{equation}
\label{eq:2ndvc1}
{\hskip 8.75mm} + A_3(\delta k_{\theta})^2 + B_3(\delta k'_{\theta})^2 + A_4(\delta k_r)^2 + B_4(\delta k'_r)^2]\dd r,
\end{equation}
where
\begin{equation}
\label{eq:abig}
A_i = \frac{\partial^2 H_1}{\partial y^2_i} - \frac{\dd}{\dd r}\big(\frac{\partial^2 H_1}{\partial y_i \partial y'_i}\big),  {\hskip 10mm} B_i = \frac{\partial^2 H_1}{\partial y'^2_i},
\end{equation}
in which $(y_i, y'_i)$ indicates the variable pairs $(m, m')$, $(k_{\rm t}, k'_{\rm t})$, $(k_{\theta}, k'_{\theta})$ or $(k_r, k'_r)$. For ideal gas, all the $A$-terms are vanishing and the $B$-terms are
\begin{equation}
\label{eq:vabsg}
B_1 = - \frac{5}{8\pi r^2 \rho_0}<0, {\hskip 2.5mm} B_2 = \frac{B_3}{2} = \frac{B_4}{2} = - \frac{\rho_0}{2\pi r^2 p^2_0}<0.
\end{equation}
These results are listed in Table~\ref{table1}.

We point out that, as shown in fig.~4 of \citet{hep10}, $\delta m$ and $\delta m'$ (or $\delta\rho$) are two different modes of perturbations. The former is the long-range (or large-scale) perturbation, whereas the latter is the short-range (or small-scale) perturbation. Similarly, $\delta k_{\theta}$, $\delta k_{\phi}$ and $\delta k_r$ are all long-range perturbations of the relevant quantities, and $\delta k'_{\theta}$, $\delta k'_{\phi}$ and $\delta k'_r$ (or $\delta p_{\theta}$, $\delta p_{\phi}$ and $\delta p_r$) are the corresponding short-range perturbations. The long- and short-range relaxation mechanisms correspond to the long- and short-range perturbations, respectively. We know that the relaxation processes for ideal gas are collisions between the particles and between the particles and the walls of the container, which are both short-range relaxations. That is the reason why all the $A$-terms, representing the long-range relaxations, vanish.

Since the $B$-terms represent short-range relaxations, and they are all negative, so the system at every local volume element takes the local maximum entropy and the system acquires its cumulative maximum entropy from all these local maxima.

\begin{table}
\caption{$A$- and $B$-terms of the second variation of entropy for both ideal gas and self-gravitating systems.}
\label{table1}
\begin{tabular}{lcc}
\hline \hline
Variables    & ~~~~~~~~~~~~~~~~Ideal gas~~~~~~~~~~~~ & Self-gravitating system \\ \hline
$(m,  m')$                  & $A_1=0$,~~~ $B_1<0$    &  $A_1>0$,~~~ $B_1<0$    \\
$(k_{\rm t}, k'_{\rm t}\ )$ & $A_2=0$,~~~ $B_2<0$    &  $A_2=0$,~~~ $B_2<0$    \\
$(k_{\theta}, k'_{\theta})$ & $A_3=0$,~~~ $B_3<0$    &  $A_3=0$,~~~ $B_3<0$    \\
$(k_r, k'_r)$               & $A_4=0$,~~~ $B_4<0$    &  $A_4=0$,~~~ $B_4<0$    \\ \hline
\end{tabular}
\end{table}

\subsection{Equilibrated self-gravitating systems}
\label{sec:ess}

The constrained entropy of equation~(\ref{eq:totcs}) is incomplete, since we just used the truncated DF of equation~(\ref{eq:3df}), and only with the second-order Jeans equation (in its virialization form) included as the additional constraint. However, we believe that the following results based on such a truncated DF are still heuristic. The second variation of the entropy of spherical self-gravitating systems is
\begin{displaymath}
\delta^2 S_{\rm t, c} =  \frac{1}{2} \int^{\infty}_0 [A_1(\delta m)^2 + B_1(\delta m')^2 + A_2(\delta k_{\rm t})^2 + B_2(\delta k'_{\rm t})^2
\end{displaymath}
\begin{displaymath}
{\hskip 13.95mm} + A_3(\delta k_{\theta})^2 + B_3(\delta k'_{\theta})^2 + A_4(\delta k_r)^2 + B_4(\delta k'_r)^2
\end{displaymath}
\begin{equation}
\label{eq:2ndvc2}
{\hskip 13.95mm} + A_5(\delta u)^2 + B_5(\delta u')^2]\dd r,
\end{equation}
where $A_i$ and $B_i$ for $(m,m')$, $(k_{\rm t}, k'_{\rm t})$, $(k_{\theta}, k'_{\theta})$ and $(k_r, k'_r)$ are calculated in the same way as those for ideal gas in equation~(\ref{eq:abig}), except that $H_1$ is replaced by $H_2$ here. The non-vanishing terms are
\begin{displaymath}
\label{eq:absg}
A_1=(\frac{\rho}{p_t}-\frac{\rho}{2 p_r}-\lambda ) \frac{G}{r^2}>0, {\hskip 2mm} B_1 = - \frac{5}{8\pi r^2 \rho}<0,
\end{displaymath}
\begin{equation}
B_2 = \frac{B_3}{2} = - \frac{\rho}{2\pi r^2 p^2_t}<0, {\hskip 5mm} B_4 = B_2 - \frac{\rho}{2\pi r^2 p^2_r}<0
\end{equation}
and
\begin{equation}
\label{eq:ab5sg}
A_5 = \frac{\partial^2 H_2}{\partial u^2} - \frac{\dd}{\dd r}\big(\frac{\partial^2 H_2} {\partial u \partial u'}\big) =0, {\hskip 5mm} B_5 = \frac{\partial^2 H_2}{\partial u'^2} = 0.
\end{equation}
From the simulation results of \citet{navarro10} and \citet{ludlow10}, we know that $p_t \leq p_r$, and with $\lambda<0$, we have $A_1>0$. These results are also listed in Table~\ref{table1}.

We see that $A_1>0$, but all the $B$-terms are negative. Hence the statistical equilibrium states of self-gravitating system are neither maximum nor minimum, but saddle-point entropy states. As analysed previously, $\delta m$ is the long-range perturbation, and $A$-terms represent long-range relaxations, whereas $\delta \rho$, $\delta p_r$ and $\delta p_t$ are short-range perturbations, and all the $B$-terms represent short-range relaxations. Thus, the saddle-point solution suggests that the equilibrium entropy is the cumulative maximum entropy from every local maximum under short-range relaxations, while simultaneously being also the global minimum entropy under long-range relaxations. Such a saddle-point solution is completely different from the case of ideal gas. This result also confirms our earlier finding of \citet{hep10}, in which we performed a preliminary study, by just using a phenomenological entropy of ideal gas of \citet{white87}.

\begin{figure}
\centerline{\includegraphics[width=0.90\columnwidth]{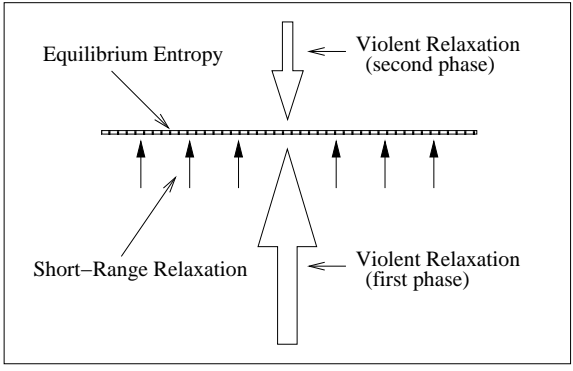}}
\caption{Illustration of the saddle-point entropy states of equilibrated self-gravitating systems and the two-phased nature of violent relaxations. The saddle-point entropy solutions are also consistent with \citet{ant62}'s proof and Binney's argument \citep{galdyn08} that the equilibrium states of self-gravitating systems are not maximum entropy states.}
\label{fig:fig1}
\end{figure}

\subsection{Free collapsing self-gravitating systems}
\label{sec:fcs}

The above discussions based on the second variations are under the presumption that the system has already settled into equilibrium states, that is, $\delta S_{\rm t, c}=0$. However, these discussions are invalid if the system has not arrived at its equilibrium states, say, is just undergoing free collapsing. In this case, the entropy variation should be analysed with its first variation.

We assume that a free-falling isolated spherical system has a given mass $M_{\rm T}$ and energy $E_{\rm T}$. In this case, the virilization relation, equation~(\ref{eq:vir1}), should not be applied here. Thus, the total conserved energy is the only constraint, given by
\begin{equation}
\label{eq:te}
E_{\rm T}= E_{\rm K} +E_{\rm V} =\int^{\infty}_0\dd r \big[(k'_{\theta} + k'_{\phi} + k'_r) - \frac{G m m'}{r}\big],
\end{equation}
and the total constrained entropy from equations~(\ref{eq:bgt}) or (\ref{eq:tots2}) for the free collapsing system (FCS) is,
\begin{displaymath}
S_{\rm FCS} =  \int^{\infty}_0 H_3(r, m, m', k_{\theta}, k'_{\theta}, k_{\phi}, k'_{\phi}, k_r, k'_r)\dd r
\end{displaymath}
\begin{displaymath}
{\hskip 7.5mm} = \int^{\infty}_0 \left\{m'\ln\left[\frac{k'^{1/2}_{\theta} k'^{1/2}_{\phi} k'^{1/2}_r}{m'^{5/2}} \right] + m' \ln(2^{7/2} \pi r^2) \right.
\end{displaymath}
\begin{equation}
\label{eq:fcss} 
{\hskip 10.9mm} + \left. \lambda(k'_{\theta} + k'_{\phi} + k'_r - \frac{G m m'} {r}) \right\} \dd r,
\end{equation}
where $H_3$ denotes all the terms enclosed in the braces, with $\lambda$ being the Lagrangian multiplier, and should always be negative\footnote{Generally, the DF derived from statistical mechanics is always a function of the energy level $\varepsilon$, as $f(\varepsilon) \sim e^{\lambda\varepsilon}$, in which we see that $f(\varepsilon) \rightarrow \infty$ as $\varepsilon\rightarrow\infty$ if $\lambda$ is positive. Hence, to protect the DF from `ultraviolet catastrophe', $\lambda$ should always be negative.}. We calculate the first variation of the entropy as
\begin{eqnarray}
\label{eq:1stvar}
\delta S_{\rm FCS} & = & \int^{\infty}_0\dd r \big[C_1\delta m + D_1 \delta m' + C_2\delta k_{\theta} + D_2 \delta k'_{\theta} \nonumber \\
 & & + C_3\delta k_{\phi} + D_3 \delta k'_{\phi} + C_4\delta k_r + D_4 \delta k'_r\big],
\end{eqnarray}
where
\begin{displaymath}
C_1=\frac{\partial H_3}{\partial m}=-\lambda\frac{Gm'}{r} > 0,
\end{displaymath}
\begin{equation}
\label{eq:l4}
C_2 = \frac{\partial H_3}{\partial k_{\theta}} = 0, {\hskip 2mm} C_3 = \frac{\partial H_3} {\partial k_{\phi}} =0, {\hskip 2mm} C_4 = \frac{\partial H_3}{\partial k_r} =0,
\end{equation}
and
\begin{displaymath}
D_1=\frac{\partial H_3}{\partial m'} = -\lambda\frac{G m}{r} + \ln\big(\frac{p^{1/2}_r p^{1/2}_{\theta} p^{1/2}_{\phi}} {\rho^{5/2}}\big) - \frac{5}{2},
\end{displaymath}
\begin{displaymath}
D_2 = \frac{\partial H_3}{\partial k'_{\theta}} = \frac{m'}{2k'_{\theta}} + \lambda, {\hskip 2mm} D_3 = \frac{\partial H_3}{\partial k'_{\phi}} = \frac{m'}{2k'_{\phi}} + \lambda,
\end{displaymath}
\begin{equation}
\label{eq:s4}
D_4 = \frac{\partial H_3}{\partial k'_r} = \frac{m'}{2k'_r}+\lambda.
\end{equation}
Note that in the free-collapsing process, especially in the starting period, the mass contained in the $r$-sphere increases dramatically and hence for any $r$, the long-range perturbation $\delta m \gg 0 $, but all the short-range perturbations, $\delta \rho$, $\delta p_{\theta}$, $\delta p_{\phi}$ and $\delta p_r$, compared with $\delta m$, are small and negligible. Thus, according to the above analysis, we have $\delta S_{\rm FCS} \approx \int \dd r C_1 \delta m>0$ and we arrive at the conclusion that the free collapsing of self-gravitating systems is an entropy-production process. The entropy production will decrease or may stop when $\delta m \sim 0$.

We put together the results of both sections~\ref{sec:ess} and \ref{sec:fcs} in Fig.~\ref{fig:fig1}. The nomenclature of two-phased violent relaxations is after \citet{soker96} who had already realized that the entropy of self-gravitating systems may not necessarily increase during the violent relaxation process\footnote{The violent relaxation we referred to in fig.~5 of \citet{hep10} should actually be the second-phase violent relaxation of this Letter.} (cf. \citealt*{tremaine86}; see also \citealt{sridhar87, kandrup87}).

\section{Discussions and Conclusions}
\label{sec:concl}

In this Letter, we investigate the stability of the statistical equilibrium of spherically symmetric collisionless self-gravitating systems by calculating the second variation of entropy. The entropy, in the Boltzmann-Gibbs form, is approximated by truncating the DF to the second-order expansion of velocity, with the constraints of the given mass, energy and the virialization relation, equation~(\ref{eq:vir1}). The main results of this Letter are as follows:
\begin{enumerate}
\item Perturbations of the relevant quantities should be divided into two categories: long-range perturbations, such as $\delta m$, and short-range perturbations, such as $\delta\rho$, $\delta p_r$ and $\delta p_t$.

\item Relaxation mechanisms of different scales should be relevant to the perturbations of the same scales. The long-range relaxations, such as violent relaxation and the long-wave Landau damping, are responsible for suppressing the long-range perturbations, whereas the short-range relaxations, such as phase-mixing, chaotic mixing or short-wave Landau damping \citep[see][]{galdyn08, mo10}, may be responsible for wiping out the short-range perturbations.

\item The statistical equilibrium states of self-gravitating systems are neither maximum nor minimum, but saddle-point entropy states. This finding conforms to our earlier result of \citet{hep10} and is also consistent with \citet{ant62}'s proof and Binney's argument \citep{galdyn08} that the equilibrium states of self-gravitating systems are not maximum entropy states.

\item Violent relaxation can be classified into two phases. The two-phased violent relaxation was first proposed by \citet{soker96} who had already realized that the entropy of self-gravitating systems may not necessarily increase during the violent relaxation process. Our saddle-point solutions support this classification.
\end{enumerate}

The free collapsing of gravitating systems, driven by (the first-phase) violent relaxation, is a rapid entropy-production process. When the long-range perturbation $\delta m$ vanishes, the violent relaxation will cease and the rapid entropy production may also stop. However, the entropy might still increase slowly, driven by the short-range relaxations to approach the final equilibrium states, while if relaxations were so efficient that the produced entropy would exceed the equilibrium entropy, then the second-phase violent relaxation might be induced to draw the system back to its equilibrium state by decreasing its entropy. Such a process, that is, oscillating around the equilibrium entropy, would last for some time with a gradually damping amplitude and finally the system would arrive at its equilibrium state.

The above description, suggested by the saddle-point solutions, may be a possible picture of the stability of the statistical equilibrium for self-gravitating systems and we further speculate that the second-phase violent relaxation may just be the long-wave Landau damping. The justification of the picture as well as the speculation need further investigations.

\section*{Acknowledgements}

This work is supported by National Basic Research Program of China, No:2010CB832805.

\label{lastpage}
\end{document}